\documentclass[aps,superscriptaddress,english,12pt,nofootinbib]{revtex4}
\DeclareUnicodeCharacter{2212}{-}
\usepackage{amssymb, amsmath, bm, dcolumn, epsf, graphicx, latexsym, slashed, simplewick}
\usepackage[caption=false]{subfig}
\usepackage[utf8,utf8x]{inputenc}
\usepackage[normalem]{ulem}
\usepackage{color}
\usepackage{float}
\usepackage{multirow}
\usepackage[toc,title,page]{appendix}

\def\be{\begin{equation}}
\def\ee{\end{equation}}
\def\bea{\begin{eqnarray}}
\def\eea{\end{eqnarray}}

\usepackage[capitalize]{cleveref}
\usepackage{comment}
\usepackage{silence}
\usepackage{subfig}
\usepackage{slashed}
\usepackage{soul}
\newcommand{\cmmnt}[1]{\ignorespaces}

\usepackage[normalem]{ulem}

\begin{document}
\title{Disentangling the Origins of the NANOGrav Signal: Early Universe Models and $\Delta N_{eff}$ Bounds}

\author{Ido Ben-Dayan}
\affiliation{Physics Department, Ariel University, Ariel 40700, Israel}
\author{Utkarsh Kumar}
\affiliation{Department of Physics, University of Ottawa, Ottawa, ON K1N6N5, Canada}
\author{Amresh Verma}
\affiliation{Physics Department, Ariel University, Ariel 40700, Israel}

\widetext

\date{\today}

\begin{abstract}
We investigate whether an Early-Universe stochastic gravitational–wave background (SGWB) can account for the common‐spectrum process reported by NANOGrav, while also being consistent with current and projected CMB measurements of extra radiation. We compute the contribution of effective number of relativistic species, $\Delta N_{eff}$, for a number of Early-Universe models proposed to explain the pulsar timing array (PTA) spectrum. We demonstrate that models predicting $\Delta N_{eff}$ above the CMB limit would be firmly excluded, implying that the NANOGrav signal in tension with these bounds must instead arise from astrophysical sources. We find that current NANOGrav 15-year dataset, sensitive up to 60 nHz, gives a negligible contribution to $\Delta N_{eff}$ and remains well below the present and future CMB detection threshold. However, when we project future PTA capabilities reaching upto 1 $\mu$Hz, even with our conservative estimate we find that Inflation, Scalar Induced Gravitational Waves (SIGW), and metastable cosmic strings can induce a $\Delta N_{eff}$ large enough for $>3.5\sigma$ detection by the Simons Observatory. 
\end{abstract}

\maketitle

\section{Introduction}
The recent detection of a common spectrum red noise, indicating a stochastic gravitational wave background {(SGWB)} in pulsar timing array (PTA) datasets, most notably in the 15-years data release by NANOGrav \cite{NANOGrav:2023gor,NANOGrav:2023hde,agazie2023nanograv,NANOGrav:2023hvm,NANOGrav:2023ctt,johnson2023nanograv,NANOGrav:2023pdq}, along with EPTA \cite{Chen:2021rqp}, PPTA \cite{Goncharov:2021oub}, and IPTA \cite{Antoniadis:2022pcn} has sparked increased interest across both the astrophysical and cosmological communities. The {Gravitational Waves} (GWs) produced by a population of inspiralling supermassive black hole binaries (SMBHBs) is the leading explanation for this observed signal \cite{NANOGrav:2023hfp}; however, the spectral properties and amplitude of the signal leave room for potential cosmological interpretations as well \cite{Ellis:2023oxs}. Several Early-Universe mechanisms, such as inflationary tensor modes \cite{Ben-Dayan:2023lwd,Vagnozzi:2023lwo,Benetti:2021uea,Vagnozzi:2022qmc,Das:2023nmm,Datta:2023vbs,Vagnozzi:2020gtf,Maiti:2024nhv}, scalar-induced GWs (SIGWs) \cite{Li:2023qua,Domenech:2024wao,Kohri:2024qpd,Chen:2025fcd,Wu:2025gwt,Wu:2024qdb,Zhou:2024yke,Zhou:2025djn}, first-order phase transitions \cite{Chen:2023bms,Gouttenoire:2023bqy,NANOGrav:2021flc,Bringmann:2023opz,Bagherian:2025puf,Guo:2023hyp}, networks of cosmic strings \cite{Datta:2024bqp, Blasi:2020mfx, Kume:2024adn, Hindmarsh:2022awe}, and non-standard scenarios like Pre Big Bang, string-gas cosmology and bouncing cosmologies \cite{Brandenberger:2011et,Ben-Dayan:2016iks,Ben-Dayan:2018ksd,Brandenberger:2020tcr,Ben-Dayan:2024aec}, can produce SGWB within the PTA frequency band.

The interpretation of the NANOGrav signal as primordial is already meaningful by just considering the primordial tensor spectrum itself. For example, the spectrum cannot be considered as a simple power law on all observable frequencies, as it is in contradiction with current LIGO bounds. Hence, there must be some break, running or any other modification of the spectrum if we wish to account for all known data sets of CMB, NANOGrav and LIGO \cite{Ben-Dayan:2019gll, Ben-Dayan:2023lwd,Jiang:2023gfe}.

Better yet, if the SGWB was cosmological in origin, it would have to be consistent not only with the observed amplitude and spectral shape of the PTA data (and all other frequencies), but also with CMB constraints. {GWs} redshifting as radiation contribute to the total energy density at recombination, altering the expansion rate. An excess in the effective number of relativistic species, $\Delta N_{eff}$, commonly parametrizes this contribution \cite{Hou_2013,Abazajian:2012ys,Ben-Dayan:2019gll}. The CMB serves as a crucial arbiter in this investigation. While future experiments such as the Simons Observatory \cite{SO:2024ntl}, CMB-S4 \cite{CMB-S4:2022pft}, and LiteBIRD \cite{LiteBIRD:2022cnt} will push these measurements to unprecedented precision, Planck's measurement of $\rm N_{eff} = 2.99 \pm 0.17$ \cite{Planck:2018vyg} already places meaningful constraints on any additional radiation components. This creates a unique opportunity for cosmological models to be contrasted with both indirect GW observations from PTAs and its imprint on the early thermal history of the Universe.

The idea of using $\Delta N_{eff}$ as a probe of the primordial tensor spectrum was thoroughly analyzed in \cite{Ben-Dayan:2019gll}. However, in the absence of a detection, the analysis heavily relayed on the UV cut-off of the spectrum. The detection of NANOGrav, and its interpretation as a primordial cosmological signal, radically alter the situation, and allows to place much stronger meaningful bounds on the current and future measured spectrum. 

In this work, we examine whether cosmological explanations for the PTA signal can be reconciled with the radiation content of the Universe as inferred from the CMB \cite{Ben-Dayan:2019gll,Nakayama:2018ptw}. We calculate the GW contribution to $\Delta N_{eff}$ for various theoretical models by integrating their spectral energy densities over the finite frequency band relevant to current and future PTAs. While present datasets probe GW frequencies up to $\sim 60$ nHz \cite{NANOGrav:2023tcn}, future PTA capabilities are expected to extend up to $1\, \mu$Hz \cite{Brazier:2019mmu,Rosado:2015epa,Moore:2014eua,Bailes:2021tot}, allowing more of the SGWB spectrum to be probed and hence increasing the corresponding $\Delta N_{eff}$. Such extrapolations of PTA sensitivity to $\mu$Hz frequencies have been investigated previously as well; for instance, \cite{Perera:2018pts} studied the SMBHB signal up to 5~$\mu$Hz for the millisecond pulsar PSR J1713$+$0747.

We focus on representative models within each class of Early-Universe GW sources. For inflationary scenarios, we explore three cases: one \texttt{standard} instant reheating scenario, a prolonged reheating scenario, and a model with running of the spectral index. We also consider scalar-induced GWs from enhanced primordial scalar spectrum with \textit{running} and \textit{running of running}, spectra arising from first-order phase transitions, and a metastable cosmic string network. For each case, we perform a Bayesian analysis combining NANOGrav likelihoods with priors informed by the relevant cosmological model. We then compute the integrated $\Delta N_{eff}$ and assess its detectability with current and future CMB experiments.

Our findings show that some models such as scalar-induced GWs lead to a large $\Delta N_{eff}$ and will likely be among the first ones to be ruled out as explanations for the PTA signal. Others, such as the standard inflationary or metastable string scenarios, yield $\Delta N_{eff}$ values that are currently undetectable but may become accessible with the next generation of CMB and PTA measurements. Phase transition models, on the other hand, produce negligible contributions to $\Delta N_{eff}$ even under optimistic extrapolations, and thus remain unconstrained by this approach.

The paper is organized as follows. In Section II, we begin by discussing various Early-Universe models and their GW spectrum contributions to $\Delta N_{\rm eff}$. The methodology and data analysis techniques are described in Section III. In Section IV, we explain the obtained results. Finally, we conclude in Section V.

\section{Physics of Early-Universe and $\Delta N_{\rm eff}$} 
Let us consider a range of Early-Universe processes that are expected to produce SGWB detectable by PTA experiments. 
Most notable cosmological sources of SGWB include inflationary scenarios, phase transitions, cosmic strings, and Scalar Induced Gravitational Waves (SIGW). We briefly review each model and its predictions. Further details can be found in the relevant references.
Following this, we will move next into the consequences of these GWs spectra for the effective number of relativistic species, $\Delta N_{\rm eff}$, and how they may play a role in Early-Universe model building.

\subsection{Inflation}
Cosmic inflation remains one of the leading theoretical frameworks explaining the {source} of the primordial gravitational wave background. Here, we give a brief theoretical outline of the mechanisms that generate GWs during the inflationary period. We then follow their evolution as they propagate through various phases of cosmic history. GWs are described as the tensor modes of the metric perturbation in the FLRW metric, which is,
\begin{equation}
    ds^2 = a^{2} (\tau)  \,\left[ d \tau^2 - \left(\delta_{ij} + h_{ij} \right) \, dx^{i}\,dx^{j} \right].
\end{equation}
Here $a$ and $\tau$ are the scale factor and the conformal time respectively, and $h_{ij}$ is the traceless-transverse symmetric $3 \times 3$ matrix describing the GW. The inflationary GWs have two polarizations $+, \times$. Each of which is governed by the following equation in Fourier space:
\begin{eqnarray}
    h_k '' + 2 \mathcal{H}\,h_k' + k^2\,h_k = 0 \label{eq: gw},
\end{eqnarray} 
where $'$ denotes a derivative with respect to conformal time, $\mathcal{H}$ is the conformal Hubble parameter and $k$ stands for the {comoving} wavenumber. For a given background one can solve eq.(\ref{eq: gw}) and calculate the primordial tensor power spectrum defined as
\begin{eqnarray}
    \mathcal{P}_{T}^{\text{prim}} &=& 8 \frac{k^{3}}{2\,\pi^{2}} |h_k|^{2} . 
\end{eqnarray}
For effective model selection, we work with the following parametrization:
\begin{eqnarray}
    \mathcal{P}_{T}^{\text{prim}} (k) = r\, A_{s} \,\left(\frac{k}{k_{*}}\right)^{n_t}.
\end{eqnarray}
 Here $r$ and $A_s$ are the tensor-to-scalar ratio and amplitude of primordial scalar power spectrum evaluated at suitable pivot scale $k_*$ that varies from one experiment to another. We fix $A_{s} = 2.1 \times 10^{-9}$ and pivot scale $ k_* = 0.05 Mpc^{-1} $  in accordance with Cosmic Microwave Background measurements from Planck \cite{Planck:2018vyg}. 
{On the other hand,} Gravitational Waves experiments such as laser-interferometer (LI) and PTA measurements report their results in terms of gravitational-wave energy spectrum measured today ($\tau = \tau_0$) as 
\begin{eqnarray}
    \Omega_{\text{GW}}^{\text{prim}} (f) &=& \frac{1}{\rho_{0}^{crit}}\,\frac{d \rho_{0}^{\text{GW}}}{d \ln f},
\end{eqnarray}
 where $f = \frac{9.72 \times  10^{-15}}{2 \,\pi \, a_0}  \frac{k}{\text{Mpc}^{-1}} \text{Hz}$ is the present-day physical frequency of a GW associated with the comoving wavenumber $k$.
 
The present-day energy spectrum, $\Omega_{\rm GW}^{\text{prim}}(f)$, of the GW generated during the inflationary epoch, is related to the primordial tensor power spectrum  $ \mathcal{P}_{T}^{\text{prim}} (f)$ via
 \begin{eqnarray}
     \Omega_{\text{GW}}^{\text{prim}} (f) &=& \frac{1}{12} \, \left[ \frac{2 \pi f}{H_0}\right]^{2} \, T_{h} (f) \, \mathcal{P}_{T}^{\text{prim}} (f),
 \end{eqnarray} 
where 
$T_{h} (f)$ is  the ``tensor transfer function" given as \cite{Boyle:2007zx},
\begin{eqnarray}
     T_{h} (f) & = & \frac{C_2 (f)\, C_3 (f) }{2 \, \left(1 + z_{re}\right)^{2}} \, \left[ \frac{\Tilde{\gamma}^{-1/2}\, 2\,\pi\,f}{\left(1 + z_{re} \right)\, H_0}\right]^{-4 / (1 + 3\,w_{re} (f))}. \,  \label{eq: tensortransfer}
 \end{eqnarray}
The final expression of the present-day GW energy spectrum for general reheating scenario is written as 
 \be
          \Omega_{\text{GW}}^{\text{prim}}(f) = \frac{r \, A_s\, C_2(f)\, C_3 (f)\, \Tilde{\gamma} }{24} \, \left(\frac{\Tilde{\gamma}^{-1/2}\,  2\,\pi\,f}{(1 + z_{re})\,  H_0}\right)^{2\,\frac{3 w_{re}(f) -1}{3 w_{re}(f) +1}}  \, \left(\frac{a_0\,H_0\,2\,\pi\,f}{k_{*}\,H_0}\right)^{n_t(f)}, \label{eq: GWInfl}
 \ee 
where the factors {$\Tilde{\gamma}$}, $C_2 (f)$ and $C_3 (f)$ are defined as 
\begin{eqnarray}
    \Tilde{\gamma} = \frac{\Omega_{m0}}{1 + z_{eq}} \frac{g_{*}(z_{re})}{g_{*}(z_{eq})}  \frac{g_{*s}^{4/3}(z_{eq})}{g_{*s}^{4/3}(z_{re})},
\end{eqnarray}
\begin{eqnarray}
       C_2 (f) = \frac{\Gamma^{2} \left( \frac{5 + 3w_{re}(f)}{2(1 + 3w_{re}(f))}\right)}{\pi } \, \left[ 1 + 2 w_{re}(f) \right]^{\frac{4}{1 + 3\, w_{re}(f)}} \,, \cr
    C_3(f) = \left[- \frac{10}{7}\frac{(98\,\Omega_{fs}^{3} - 589\,
    \Omega_{fs}^{2} + 9380\,\Omega_{fs}- 55125)}
  {(15 + 4\,\Omega_{fs})(50 + 4\,\Omega_{fs})(105 + 4\,\Omega_{fs}^{})}\right]^{2},
\end{eqnarray}
 with $\Omega_{m0}$ denoting the relative present-day matter energy density, $\Omega_{fs} \equiv \frac{\rho^{fs}}{\rho^{crit}} $ is the fraction of critical density which is free-streaming relativistically (e.g., neutrinos) at redshift $z_{re}$, and $g_{*}(z), \, g_{*s}(z)$ are the number of effective relativistic degrees of freedom at redshift $z$, as measured by the energy density $\rho(z)$ or the
entropy density $s(z)$, respectively \cite{Boyle:2007zx}. 
In the above expressions, $w_{re}$ and $z_{re}$ are the reheating equation of state (eos) and reheating redshift, i.e., $z$ at which the Universe becomes radiation-dominated. We consider 3 different models for the Early Universe:

\noindent \texttt{Reheating:} First, we consider a finite reheating era, with a constant $w_{re}$, after the end of inflation and before the onset of radiation domination. For this case, in the expression of $\Omega_{GW}$, we can replace parameter $1+z_{re}$ with $T_{re}$ in GeV, i.e., the temperature at the end of the reheating era using the following relation:
\begin{eqnarray}
    \frac{1+z_{re}}{1+z_{eq}} &=& \frac{T_{re}}{T_{eq}} \nonumber \\
    1+ z_{re} & \approx & \frac{3400 \times 10^{11}}{80} \frac{T_{re}}{GeV}.
\end{eqnarray}
where we have used the numerical value of $z_{eq}$ and $T_{eq}$ from Planck 2018 data \cite{Planck:2018vyg}.

\noindent \texttt{Standard:} Next, we consider the standard instantaneous reheating scenario meaning the Universe enters the radiation-dominated era immediately after the end of inflation. Therefore, the present-day primordial GW energy spectrum reduces to
\begin{eqnarray}
    \Omega_{\text{GW}}^{\text{prim,i}}(f) = \frac{r \, A_s\, \Tilde{\gamma} }{24}  \, \left(\frac{a_0\,H_0\,2\,\pi\,f}{k_{*}\,H_0}\right)^{n_t}, \label{eq: instant}
\end{eqnarray}

\noindent \texttt{Running:} Finally, we consider a frequency-dependent tilt of the spectrum, in such scenario the tensor spectral index $n_t$ takes the following form:
\begin{equation}
    n_t(f) = n_t + \frac{\alpha_t}{2} \log\left( \frac{f}{f_{yr}}\right),
\end{equation}
with the running of the spectrum for tensor perturbations $\alpha_t$ is defined as 
\begin{equation}
    \alpha_t = \dfrac{d^2\log \mathcal{P}^{\text{prim}}_T}{d\log f^2}|_{f=f_*},
    \label{eq:running}
\end{equation}
 $f_{\text{yr}}$ being the frequency mode corresponding to the NANOGrav data ($f_{\rm yr} = 1 \rm yr^{-1} \sim 3.17 \times 10^{-8}$\,Hz).

\subsection{Scalar Induced Gravitational Waves}
In general it is of utmost interest to directly or indirectly measure the primordial scalar power spectrum on the largest possible frequency range to discern between models of the Early-Universe \cite{Ben-Dayan:2009fyj,Bringmann:2011ut,Chluba:2012we,Ben-Dayan:2013eza,Ben-Dayan:2014iya, Ben-Dayan:2015zha,Ben-Dayan:2013fva,Ben-Dayan:2016iks, Ben-Dayan:2018ksd, Artymowski:2020pci,Ben-Dayan:2023rlj}. The recent measurements of GW at various frequencies, allow further constraints due to the phenomenon of scalar induced GW. The decomposition theorem guarantees that, at linear order in perturbation, scalar, vector, and tensor perturbations evolve independently, with each evolution governed by its corresponding equation of motion \cite{Mollerach:2003nq, Baumann:2007zm,Ananda:2006af,Kumar:2024bvp}. However, at second order in perturbations, an extra source term is generated for the tensor perturbations due to the mode coupling of scalar metric
fluctuations, which unavoidably leads to a stochastic GW background in the form of SIGW. In the PTA band, the spectral density of the SIGW is as follows \cite{Ben-Dayan:2019gll,Domenech:2021ztg,Pi:2020otn}:
\begin{eqnarray}
    \Omega_{\rm GW} =\Omega_r \left(\frac{g_{*}(f)}{g_{*}^{\rm 0}}\right) \left(\frac{g_{*,s}^{\rm 0}}{g_{*,s}(f)}\right)^{-4/3} \overline{\Omega}_{\rm GW}\,,
\end{eqnarray}
where $\Omega_{ r}/g_*^0 \simeq 2.72 \times 10^{-5}$ is radiation energy density per relativistic degree of freedom, in units of the critical density, $g_{*,s}^0 \simeq 3.93$ is the effective number of relativistic degrees of freedom contributing to the radiation entropy today, and $g_*(f)$ and $g_{*,s}(f)$ denote the effective numbers of relativistic degrees of freedom in the Early-Universe that contribute to the radiation energy density and entropy density respectively. The last factor denotes the GW spectrum at the time of production, which for $w=1/3$ and constant $c_s$ is given by \cite{Domenech:2021ztg}
\begin{eqnarray}
    \overline{\Omega}_{\rm GW}=\int_0^\infty dv\int_{|1-v|}^{1+v}du\,{\cal T}(u,v,c_s,w=1/3){{\cal P}_{\cal R}(ku)}{{\cal P}_{\cal R}(kv)}\,,
    \label{eq: SIGW}
\end{eqnarray}
where ${\cal P}_{\cal R}(k)$ is the primordial spectrum of curvature fluctuations, the transfer function is given by
\begin{align}
    {\cal T}(u,v,c_s,w=1/3) =& \, \frac{y^2}{3c_s^4}\left(\frac{4v^2-(1-u^2+v^2)^2}{4u^2v^2}\right)^2\nonumber\\&\times
\left\{\frac{\pi^2}{4}y^2\Theta[c_s(u+v)-1]
+\left(1-\frac{1}{2}y \ln\left|\frac{1+y}{1-y}\right|\right)^2\right\}\,,
\end{align}
and
\begin{eqnarray}
    y=\frac{u^2+v^2-c_s^{-2}}{2 uv}\,.
\end{eqnarray}
These formulae agree with the ones in Refs. \cite{Espinosa:2018eve,Kohri:2018awv} for $c_s^2=1/3$. Equation (\ref{eq: SIGW}) clearly illustrates the dependence of the SIGW spectrum on the scalar spectrum input, namely, $\Omega_{\rm GW} \propto {\cal P}_{\cal R}^2$. The most common parametrization for the primordial scalar spectrum with constant tilt ($n_s$) is strictly constrained by CMB measurements and the resulting GW amplitude cannot reach the PTA detectable magnitude. The same is true for the immediate extension of this spectrum with \textit{running} ($\alpha$), as reported recently by the ACT Collaboration. Therefore, we consider the following form for the scalar spectrum with \textit{running} ($\alpha$) and \textit{running of running} ($\beta$):
\begin{eqnarray}
    {\cal P}_{\cal R}(k) = A_s \left(\frac{k}{k_0}\right)^{n_s(k_0)-1+\frac{\alpha(k_0)}{2} \text{ln} \frac{k}{k_0}+\frac{\beta(k_0)}{6} \text{ln}^2\frac{k}{k_0}},
\end{eqnarray}
where $A_s$ is the amplitude of the scalar spectrum. The amplitude ($A_s$) and tilt ($n_s$) are set to the maximal likelihood value of Planck, i.e., $A_s = 2.1 \times 10^{-9}$, $n_s = 0.96$, and $k_0 = 0.05 \, \rm Mpc^{-1}$ is the chosen pivot scale.

\subsection{Phase Transition}
The quantum and thermal fluctuations of the scalar field $\phi$ typically the Higgs field \cite{Barger:2007im,Damgaard:2015con}, though in models with additional scalars it may be another field \cite{Cline:1996mga,FileviezPerez:2008bj,Dorsch:2013wja} trigger phase transitions that tunnel through or fluctuate across a potential barrier.
This causes the formation of bubbles where the scalar field settles into the true vacuum state. When these bubbles become larger in size, they propagate through the surrounding plasma, still occupied by the scalar field in its false vacuum \cite{Lewicki:2020azd}. The expansion and collision of these bubbles, along with the sound waves generated within the plasma, can generate a signature of the SGWB \cite{Cutting:2018tjt}.

When the bubble walls interact with the surrounding plasma, the majority of the energy released during the phase transition is typically converted into plasma motion, leading to the dominance of sound waves in the resulting SGWB \cite{Ellis:2018mja}. In some models, where the interaction between the bubble walls and the plasma is absent or minimal, or in cases where the energy released during the phase transition is substantial enough that the plasma's friction is insufficient to halt the bubble walls' acceleration (the ``runaway scenario"), SGWB is predominantly generated as a result of Bubble collisions \cite{Ellis:2020awk}.

The SGWB spectrum sourced by bubbles (referred to as \texttt{Bubble}) and sound waves (referred to as \texttt{Sound}) can be written in terms of these parameters as
\begin{align}
\Omega_b(f)&=\mathcal{D}\,\tilde{\Omega}_b\left(\frac{\alpha_*}{1+\alpha_*}\right)^2\left(H_*R_*\right)^2\mathcal{S}(f/f_{b})\label{eq:omega_pt_bubble},\\
    \Omega_s(f)&=\mathcal{D}\,\tilde{\Omega}_s\Upsilon(\tau_{\rm sw})\left(\frac{\kappa_s\,\alpha_*}{1+\alpha_*}\right)^2\left(H_*R_*\right)\mathcal{S}(f/f_{s})\label{eq:omega_pt_sound} \,.
\end{align}
Here $\tilde{\Omega}_b=0.0049$ \cite{Huang:2017laj} and $\tilde{\Omega}_s=0.036$ \cite{Hindmarsh:2017gnf}; the efficiency factor $\kappa_s=\alpha_*/(0.73+0.083\sqrt{\alpha_*}+\alpha_*)$ \cite{Espinosa:2010hh} gives the fraction of the released energy that is transferred to plasma motion in the form of sound waves, $\alpha_*$ is the strength of transition, $H_*R_*=R_*/H_*^{-1}$ is the average bubble separation at percolation, $H_*^{-1}$, in units of Hubble radius, and $\mathcal{D}$ accounts for the redshift of the GW energy density \cite{Ellis:2020awk},
\begin{equation}\label{eq:dilution}
    \mathcal{D} = \frac{\pi^2}{90}\frac{T_0^4}{M_{\scriptscriptstyle \rm Pl}^2H_0^2}\,g_{*}\left(\frac{g_{*,s}^{\rm eq}}{g_{*,s}}\right)^{4/3}\simeq 1.67\times 10^{-5} \,.
\end{equation}
We recall that $T_0$ and $H_0$ denote the photon temperature and Hubble rate today. The degrees of freedom $g_*$ and $g_{*,s}$ in Eq.~\eqref{eq:dilution} are evaluated at $T = T_*$, and $g_{*,s}^{\rm eq}$ is the number of degrees of freedom contributing to the radiation entropy at the time of matter-radiation equality.
The production of GWs from sound waves stops after a period $\tau_{\rm sw}$, when the plasma motion turns turbulent \cite{Guo:2020grp}. In Eq.~\eqref{eq:omega_pt_sound}, this effect is taken into account by the suppression factor 
\begin{equation}
\Upsilon(\tau_{\rm sw}) = 1 - (1+2\tau_{\rm sw}H_*)^{-1/2},
\end{equation}
where the shock formation time scale, $\tau_{\rm sw}$, can be written in terms of the phase transition parameters as $\tau_{\rm sw}\approx R_*/\bar U_f$, with $\bar U^2_f\approx3\kappa_s\alpha_*/[4(1+\alpha_*)]$ \cite{Weir:2017wfa}. The functions $\mathcal{S}_{b,s}$ describe the spectral shape of the SGWB and are expected to peak at the frequencies 
\begin{equation}\label{eq:pt_peak_f}
    f_{b,s}\simeq 48.5\,{\rm nHz}\; g_*^{1/2}\bigg(\frac{g_{*,s}^{\rm eq}}{g_{*,s}}\bigg)^{1/3}\bigg(\frac{T_*}{1\,\rm GeV}\bigg)\:\frac{f_{b,s}^*R_*}{H_*R_*} \,,
\end{equation}
where the values of the peak frequencies at the time of GW emission are given by $f_b^*=0.58/R_*$ \cite{Jinno:2016vai} and $f_s^*=1.58/R_*$ \cite{Hindmarsh:2017gnf}. 

The shape of the spectral functions can usually be approximated with a broken power law of the form \cite{Hindmarsh:2020hop}
\begin{equation}\label{eq:pt_spec_shape}
    \mathcal{S}(x)=\frac{1}{\mathcal{N}}\frac{(a+b)^c}{(bx^{-a/c}+ax^{b/c})^c} \,.
\end{equation}
Here $a$ and $b$ are two real and positive numbers that give the slope of the spectrum at low and high frequencies, respectively; $c$ parametrizes the width of the peak. The normalization constant, $\mathcal{N}$, ensures that the logarithmic integral of $\mathcal{S}$ is normalized to unity and is given by \cite{Hindmarsh:2020hop}
\begin{equation}
    \mathcal{N}=\left(\frac{b}{a}\right)^{a/n}\bigg(\frac{nc}{b}\bigg)^c \:\frac{\Gamma\left(a/n\right)\,\Gamma\left(b/n\right)}{n\,\Gamma(c)} \,,
\end{equation}
where $n=(a+b)/c$ and $\Gamma$ denotes the gamma function.

\subsection{Metastable Cosmic Strings}

Cosmological phase transitions in the Early Universe also give rise to one-dimensional topological defects called cosmic strings. Global U(1) symmetry breaking, as in axion models, leads to global-string networks whose gravitational wave signals are suppressed due to energy loss via light pseudo-Nambu–Goldstone bosons (“axions”) \cite{Cheng:2024axj}. We focus instead on cosmic strings arising from the spontaneous breaking of a local U(1) symmetry within a grand unified theory (GUT) framework~\cite{Vilenkin:2000jqa,Dvali:1994ms, Jeannerot:2003qv}.

Metastable cosmic strings can decay via quantum tunneling into segments connecting monopole–antimonopole pairs. In the semiclassical approximation, the decay rate per unit length is given by~\cite{Vilenkin:1982hm, Preskill:1992ck, Leblond:2009fq, Monin:2008mp}
\begin{eqnarray} \label{eq: decayrate}
    \Gamma_d = \frac{\mu}{2\pi}\,\exp\left(-\pi\kappa\right) \,,\quad \kappa = \frac{m^2}{\mu} \,,
\end{eqnarray}
where $m$ is the monopole mass and $\mu$ is the string tension. The spectral energy density of the SGWB sourced by a metastable cosmic-string network is given as \cite{Buchmuller:2021mbb}:
\begin{eqnarray}\label{eq: omegacs}
    \Omega_{\rm GW}\left(f\right) = \frac{16\pi\,\left(G\mu\right)^2}{3H_0^2f}\sum_{k} k \, P_k \int_{0}^{z_s} \frac{dz^\prime}{H(z^\prime)(1+z^\prime)} \, n(2k/f^\prime, t(z^\prime)) \, ,
\end{eqnarray}
where $f^\prime = 2k/\ell$ indicates the GW frequency emitted by a loop of length $\ell$ oscillating in its $k$th harmonic excitation, $z_s$ is redshift when metastable strings behave as stable strings, $t(z^\prime)$ is the time of GW emission and $P_k = \frac{\Gamma}{k^{4/3}\zeta(4/3)}$ with $\Gamma \simeq 50$ being the power emitted by a single loop. We use the loop number densities $n(\ell, t)$ from \cite{Buchmuller:2021mbb}, evaluated at relevant cosmological epochs, to compute the GW spectrum in this work.

\subsection{$\bm{N}_{\text{eff}}$}
So far, we have described multiple Early-Universe mechanisms that can give rise to a detectable SGWB explaining the PTA measurements. If this interpretation is correct, it would provide a direct window into physics at energy scales far beyond the reach of terrestrial experiments. In particular, a primordial SGWB would have profound implications for the thermal history of the Universe, including the number of light, relativistic species present in the Early Universe. This is typically quantified through the parameter $\Delta N_{\rm eff}$, which measures deviations from the Standard Model prediction for the effective number of relativistic degrees of freedom. The gravitational wave contribution to the number of relativistic degrees of freedom is given by:
\begin{equation}
\label{eq:Ngw-rho}
N_{\rm eff}^{\rm GW} = \frac{8}{7}\frac{\rho_{GW}}{\rho_\gamma}\,\,.
\end{equation}
From $T\gtrsim1$ MeV to the present time $\rho_{GW}$ scales as $a^{-4}$, while the photon energy density evolves as $\rho_\gamma\sim 1/(a^4 g_{*,s}^{4/3})$ (i.e., by keeping entropy constant). The result is:
\begin{equation}
\label{eq:Ngw-rho-gstar}
N_{\rm eff}^{\rm GW} = \frac{8}{7}\bigg(\frac{g_{*,s}(T\gtrsim1\,\text{MeV})}{g_{*,s}(T_0)}\bigg)^{\frac{4}{3}}\frac{\rho_{GW}}{\rho_\gamma}\bigg|_{\eta = \eta_0}\,\,,
\end{equation}
where $T_0$ is the temperature of the CMB at the present time $\tau = \tau_0$. Multiplying both $\rho_{GW}$ and $\rho_\gamma$ by $1/\rho_c$, one obtains:
\begin{equation}
\label{eq:Ngw-rho-gstar-f}
N_{\rm eff}^{\rm GW} = \frac{8}{7}\bigg(\frac{g_{*,s}(T\gtrsim1\,\text{MeV})}{g_{*,s}(T_0)}\bigg)^{\frac{4}{3}}\frac{\rho_c}{\rho_\gamma}\int d \log f\,\Omega_{GW}(f)\,\,.
\end{equation}
Inserting numerical values one can find \cite{Maggiore:1999vm}:

\begin{eqnarray}
N_{\rm eff}^{\rm GW} &\approx& \frac{h^2}{{5.6 \times 10^{-6}}}\int d \log f\,\Omega_{GW}(f) \nonumber \\
 &\approx& 1.8 \times 10^{5}\, \int df \frac{\Omega_{GW} (f) \, h^2}{f} .
\end{eqnarray} 
Therefore,
\begin{eqnarray}
    \Delta N_{eff}\approx 1.8 \times 10^{5}\, \int_{f_{min}}^{f_{max}} df \frac{\Omega_{GW} (f) \, h^2}{f}\label{eq:Neff},
\end{eqnarray} 
where the values of $f_{min}$ and $f_{max}$ depend on the epoch of interest and the maximum temperature reached in the Big Bang era. 

\section{Methodology and data analysis} 
\begin{figure}[h!]
    \centering
    \includegraphics[width=\textwidth, height=0.65\textwidth]{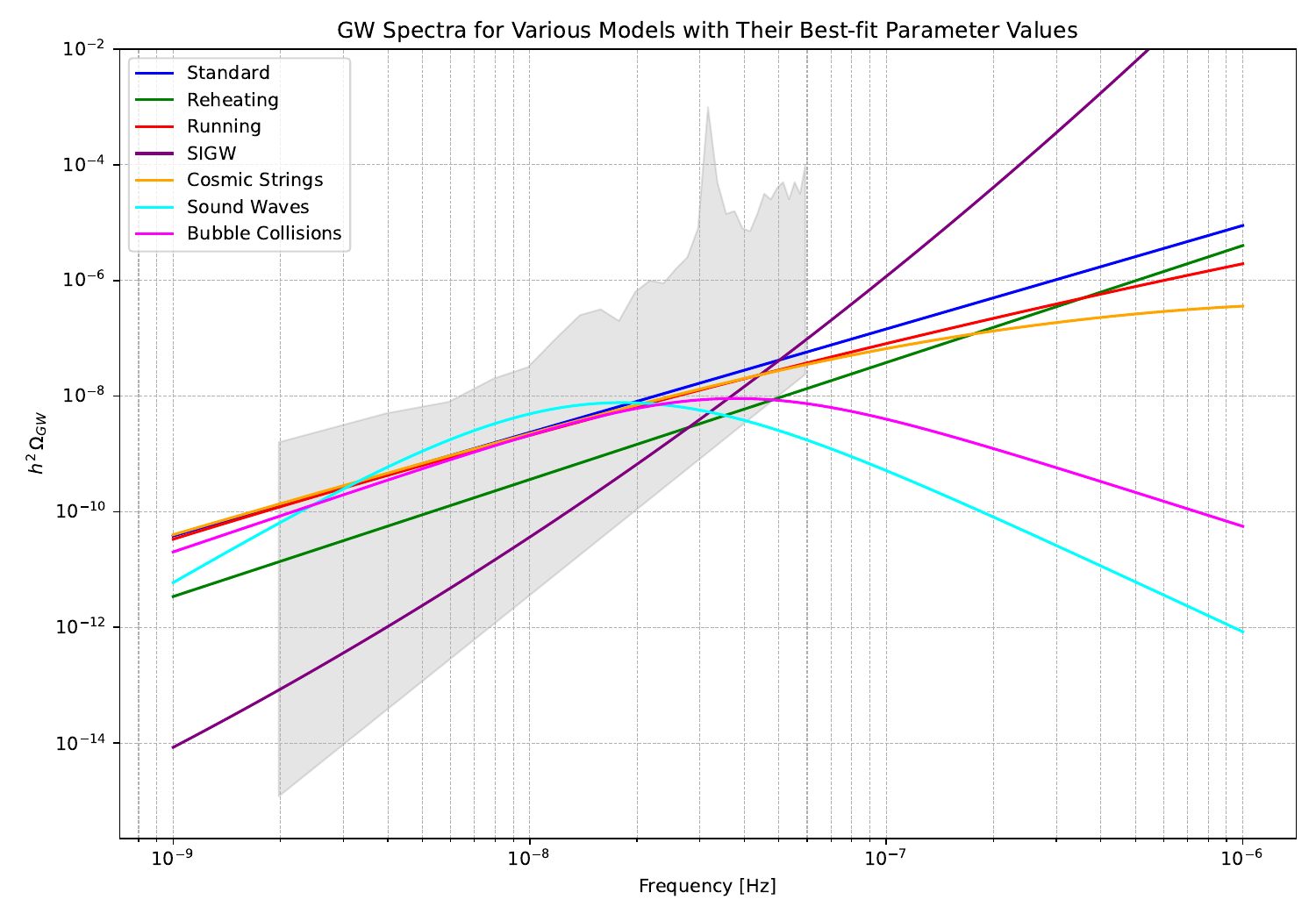}
    \caption{The present-day GW spectrum of various Early-Universe models taking the best-fit values of their parameters. The shaded region shows NANOGrav sensitivity.}
    \label{fig: omgw}
\end{figure}

We perform a Bayesian analysis using the publicly available python wrapper \texttt{PTArcade} \cite{Mitridate:2023oar} in ``ceffyl" \cite{Lamb:2023jls} mode. We adopt the flat prior on free parameters as given in \cref{tab:np_priors}.
We use \texttt{getdist} to analyze and plot the chains. We define $\log_{10} \Delta N_{\rm eff}$ as a derived parameter using \texttt{getdist} to find constraints on it, where $\Delta N_{\rm eff}$ is defined in Eq. \ref{eq:Neff}. To avoid the pulsar-intrinsic excess noise at high frequencies, we consider the first 14 frequency bins as signals for the GW background. 
We find the constraints on $\Delta N_{eff}$ against the 15-yr NANOGrav measurements integrating Eq. \ref{eq:Neff} to the same frequency range as that of the NANOGrav observations, i.e., $f \in [2 \times 10^{-9}, 6 \times 10^{-8}]Hz$. NANOGrav has over $T \sim$ 15 years of measurement data, and each pulsar is observed every $\Delta t \sim$ 1 - 3 weeks. Hence, it is most sensitive in the frequency range of $\sim$ 2 nHz to 1 $\mu$Hz (i.e. $1/T < f < 1/2\Delta t$ ) \cite{Brazier:2019mmu}. Therefore, for the forecast, we extrapolate the parameter $\log_{10} \Delta N_{eff}$ for various $f_{max}$ taking $f_{max} \in [3 \times 10^{-9}, 1 \times 10^{-6}]Hz$. We use this constraint to assess the combined ability of future PTA and CMB experiments to rule out various cosmological models proposed to explain the SGWB signal observed in the nHz range. The Simon Observatory is expected to reach a precision of $\sigma(N_{\rm eff})= 0.045$, while the CMB-S4 is expected to reach $\sigma(N_{\rm eff})= 0.027$. 
The current limit from \texttt{Planck}, however, is $\sigma(N_{\rm eff})= 0.19$ (TT+TE+EE+lowE+lensing).
To get a feeling of the expected contribution of SGWB to $N_{eff}$ we plot the predicted present day GW relative energy density $h^2 \Omega_{GW}$ of the different models using their Best-fit values in Figure \ref{fig: omgw}. These values are given in the text and in \cref{tab: allmodel}. Obviously as the relative energy density of GW increases we expect a larger contribution to $N_{eff}$.
\begin{table}[H] \label{tab:np_priors_filtered}
\centering
\caption{Prior distributions for Early Universe models discussed in this report.}
\label{tab:np_priors}
\small
\begin{minipage}[t]{0.49\textwidth}
\centering
\begin{tabular}{@{}c@{\hspace{6pt}}c@{}}
\hline
\textbf{Parameter} & \textbf{Prior} \\
\hline
\multicolumn{2}{c}{\textbf{Inflation}} \\
$r$ & log-uniform $[-26, -1.44]$ \\
$n_t$ & uniform $[0, 10]$ \\
$w$ & uniform $[-\frac{1}{3}, 1]$ \\
$\alpha_t$ & uniform $[-2, 2]$ \\
$T_{\rm rh}$ [GeV] & log-uniform $[-5, 15]$ \\
\hline
\multicolumn{2}{c}{\textbf{Scalar-induced GWs}} \\
$\alpha$ & uniform $[-0.011, 0.037]$ \\
$\beta$ & uniform $[-0.002, 0.046]$ \\
\hline
\multicolumn{2}{c}{\textbf{Metastable cosmic strings}} \\
$G\mu$ & log-uniform $[-10, -3]$ \\
$\sqrt{\kappa}$ & uniform $[7, 9]$ \\
\hline
\end{tabular}
\end{minipage}
\hfill
\begin{minipage}[t]{0.49\textwidth}
\centering
\begin{tabular}{@{}c@{\hspace{6pt}}c@{}}
\hline
\textbf{Parameter} & \textbf{Prior} \\
\hline
\multicolumn{2}{c}{\textbf{Phase transition}} \\
$T_*$ [GeV] & log-uniform $[-4, 4]$ \\
$\alpha_*$ & log-uniform $[-2, 1]$ \\
$H_*R_*$ & log-uniform $[-3, 0]$ \\
$a$ (Bubble) & uniform $[1, 3]$ \\
$a$ (Sound) & uniform $[3, 5]$ \\
$b$ (Bubble) & uniform $[1, 3]$ \\
$b$ (Sound) & uniform $[2, 4]$ \\
$c$ (Bubble) & uniform $[1, 3]$ \\
$c$ (Sound) & uniform $[3, 5]$ \\
\hline
\end{tabular}
\end{minipage}
\end{table}

\section{Results and Analysis} 

\begin{figure}[h!]
    \centering
    \begin{minipage}{\linewidth}
        \centering
        \includegraphics[width=\textwidth, height=0.5\textwidth]{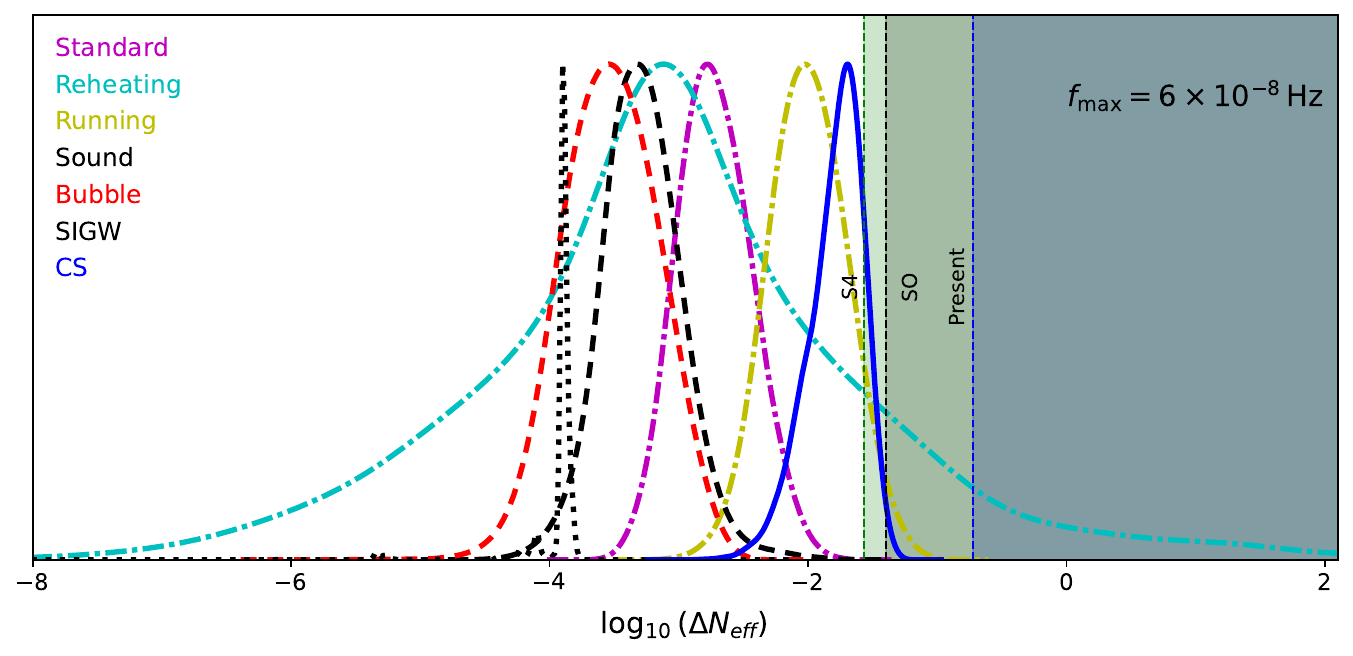} 
    \end{minipage}

    \begin{minipage}{\linewidth}
        \centering
        \includegraphics[width=\textwidth, height=0.5\textwidth]{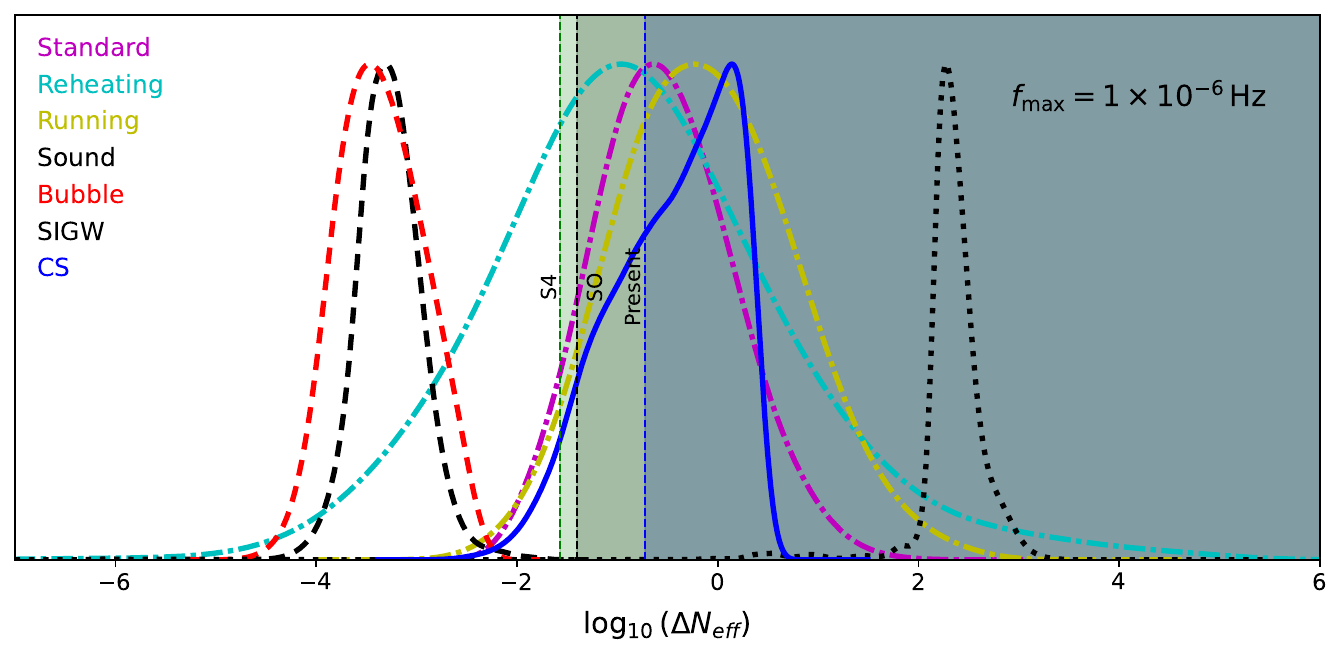} 
        \caption{The marginalized 1D posteriors on $\log_{10} \Delta N_{eff}$ for all models first calculated only for NANOGrav probed frequency range (Top) and then extrapolated till $f_{max} = 1 \, \rm \mu Hz$ (Down). For reference, we also present the `detectability zone' from (present) Planck 2018, SO, and CMB-S4 experiments, corresponding to the dashed vertical lines at $\sigma_{P18}= 0.19$, $ \sigma_{SO}= 0.045$, and $ \sigma_{S4}= 0.027$ respectively.}
        \label{fig:Newneff}
    \end{minipage} 
\end{figure}

Figure~\ref{fig:Newneff} shows marginalized 1D posteriors on $\rm log_{10}{\Delta N_{eff}}$ for different models, compared against the detectability threshold of CMB experiments in their ability to constrain $\Delta N_{eff}$. The inferred 68\% constraints on the all model parameter discussed in this work from the NANOGrav 15-year data is reported in \cref{tab: allmodel}, where the derived quantity, $\log_{10} \Delta N_{eff}$, is calculated by integrating up to maximum present NANOGrav sensitivity, i.e., $f_{max}$ = 60 nHz. In \cref{tab: significance}, we summarize the inferred 68\% constraints on $\rm log_{10}{\Delta N_{eff}}$ with `detectability significance' for our current CMB threshold, $\sigma_{\textbf{present}}$ and that for SO, $\sigma_{\textbf{SO}}$. The posterior on $\rm log_{10}{\Delta N_{eff}}$ exhibit a long tail towards large $\rm \Delta N_{eff}$. Hence, to be conservative in our estimate we report the `detectability significance' (= $\frac{\rm \Delta N_{eff}}{\sigma_{\rm N_{eff}}}$) with just the median value of $\rm \Delta N_{eff}$ obtained.

\begin{table}[ht] 
\centering
\begin{tabular}{|l|c|c|c|c|c|c|}
\hline
\textbf{Model} & ${n_t}$ & $\rm log_{10}{r}$ & ${\alpha_t}$ & ${\rm log_{10}T_{\mathrm{rh}}}$ (GeV) & ${w_{re}}$ & $\log_{10} \Delta N_{eff}$ \\
\hline
\texttt{Standard}   & $1.79_{-0.38}^{+0.33}$ & $-8.3_{-2.6}^{+3.0}$ & ---     & ---  & --- &  $-2.72_{-0.32}^{+0.27}$ \\
\hline
\texttt{Running}     & $2.48_{-0.56}^{+1.20}$ & $-14.4_{-11.0}^{+4.7}$ & $-0.09_{-0.12}^{+0.10}$ & ---  & --- &  $-2.00 \pm 0.31$  \\
\hline
\texttt{Reheating}   & $1.77_{-0.75}^{+0.33}$ & $-8.5_{-2.0}^{+6.9}$ & --- & $3.4_{-8.1}^{+3.1}$& $0.43_{-0.29}^{+0.17}$ & $-3.1 \pm 1.5$\\
\hline
\end{tabular}

\vspace{1em}

\begin{tabular}{|l|c|c|c|c|c|}
\hline
\textbf{Model} & $\rm log_{10}(G\mu)$ & $\sqrt{\kappa}$ & $\alpha$ & $\beta$ & $\log_{10} \Delta N_{eff}$ \\
\hline
\texttt{CS}   & $-5.44^{+0.76}_{-0.39}$ & $7.887^{+0.039}_{-0.11}$ & --- & --- & $-1.78^{+0.098}_{-0.210}$ \\
\hline
\texttt{SIGW}  & --- & --- & $0.0281^{+0.0087}_{-0.0011}$ & $0.00995^{+0.00076}_{-0.00088}$ & $-3.894_{-0.035}^{+0.031}$ \\
\hline
\end{tabular}

\vspace{1em}

\begin{tabular}{|l|c|c|c|c|}
\hline
\textbf{Model} & $\rm log_{10}{\alpha}$ & $\rm log_{10}{T_*}$ (GeV) & $\rm log_{10}{H_* R_*}$ & $\log_{10} \Delta N_{eff}$ \\
\hline
\texttt{Bubble}   & $0.29^{+0.65}_{-0.29}$ & $-0.71^{+0.36}_{-0.47}$ & $-0.36^{+0.32}_{-0.16}$ &  $-3.53 \pm 0.38$ \\
\hline
\texttt{Sound}  & $0.24^{+0.46}_{-0.56}$ & $-1.71^{+0.38}_{-0.43}$ & $-0.71^{+0.26}_{-0.43}$ &  $-3.28^{+0.28}_{-0.32}$ \\
\hline
\end{tabular}

\caption{The inferred 68\% constraints on the all model parameter discussed in this work from the NANOGrav 15-year data. The derived quantity, $\log_{10} \Delta N_{eff}$, is calculated by integrating up to maximum present NANOGrav sensitivity, i.e., $f_{max}$ = 60 nHz.}
\label{tab: allmodel}
\end{table}

We begin with the discussion on the inflationary scenario. From the reconstructed posterior distribution for the \texttt{Standard} model parameters $n_t$ and $\log_{10}r$ together with the derived quantity $\log_{10} \Delta N_{eff}$ we find that $\log_{10} \Delta N_{eff}$ is correlated with the tilt of the spectrum, $n_t$, and anti-correlated with the tensor-to-scalar ratio, $\log_{10} r$. The 1$\sigma$ constraints obtained for the parameters are $n_t = 1.79_{-0.38}^{+0.33}$, $\log_{10}r = -8.3_{-2.6}^{+3.0}$, and $\log_{10} \Delta N_{eff} = -2.72_{-0.32}^{+0.27}$. And the obtained 1$\sigma$ constraints for the \texttt{Reheating} and \texttt{Running} cases are $\log_{10}{T_{re}} = 3.4_{-8.1}^{+3.1}$, $\log_{10}r = -8.5_{-2.0}^{+6.9}$, $w_{re} = 0.43_{-0.29}^{+0.17}$, $n_t = 1.77_{-0.75}^{+0.33}$, $\log_{10} \Delta N_{eff} = -3.1 \pm 1.5$ and $n_t = 2.48_{-0.56}^{+1.20}$, $\log_{10}r = -14.4_{-11.0}^{+4.7}$, $\alpha_t = -0.09_{-0.12}^{+0.10}$, $\log_{10} \Delta N_{eff} = -2.00 \pm 0.31$, respectively. The 1$\sigma$ uncertainty on $\log_{10} \Delta N_{eff}$ for the \texttt{Reheating} case is considerably larger than the other two cases. This is because we have more free parameters in the \texttt{Reheating} model. Preserving the same constraints on the parameters of all three models we extrapolate the integral for $N_{eff}$ for higher $f_{max}$ values and plot the 68 \% CL, result is shown in Fig. 2. 
With current CMB limits, we do not achieve a smoking-gun detection for the \texttt{Standard} case, reaching only about a $\sim 1.5\sigma$ significance. However, the Simons Observatory (SO) is expected to provide a $>6\sigma$ detection. The contribution from the \texttt{Reheating} scenario remains consistent with noise under present limits at $\sim 0.8\sigma$, but yields a $>3.5\sigma$ detection with SO. The best detectability is obtained for the \texttt{Running} model, with $\sim 4\sigma$ and $\sim 16\sigma$ significance under current and SO limits, respectively.

In the SIGW model, for the Bayesian analysis, we take a flat prior on $\alpha \in [-0.011, 0.037]$ and $\beta \in [-0.002, 0.046]$, respecting the 2$\sigma$ bound set by Planck 2018 \cite{Planck:2018jri}. This results in the constraints on the \textit{running}, $\alpha = 0.0281^{+0.0087}_{-0.0011}$, the \textit{running of running}, $\beta = 0.00995^{+0.00076}_{-0.00088}$, at 68\% CL. The 1$\sigma$ constraint on the derived parameter $\Delta N_{eff}$ integrated until 60 nHz is  $\log_{10} \Delta N_{eff} = -3.894_{-0.035}^{+0.031}$. The extrapolated integral for $f_{max} = \rm 1 \, \mu Hz$ is plotted at the 68 \% CL, shown in Fig. 2. 
This extrapolation constrains $\Delta N_{eff}$ to be $\log_{10} \Delta N_{eff} = 2.336_{-0.153}^{+0.231}$. This is a huge contribution to the $N_{eff}$ and it will be detected by over 1000$\sigma$ by any CMB experiment. This means we will be able to exclude the SIGW model as an explanation for PTA signal much before it reaches the sensitivity of 1$\mu$Hz. 

\begin{figure}
    \centering
    \includegraphics[width=0.8\linewidth]{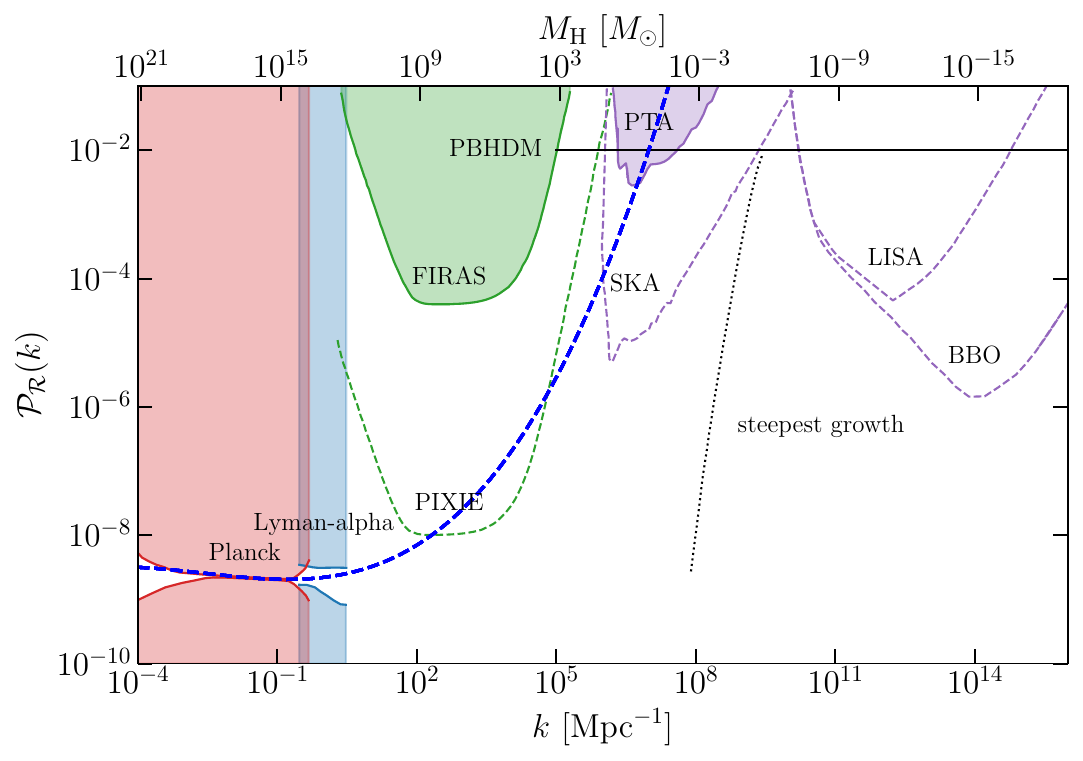}
    \caption{The primordial scalar power spectrum as a function of wavenumber (lower x-axis) and black hole mass (upper x-axis). Current constraints are given by shaded regions, and expected future ones by dashed lines, with the name of each experiment. The blue dashed line is the scalar spectrum using the best-fit values that generate the observed NANOGrav signal as a result of SIGW. It is clear that the spectrum exceeds the $10^{-2}$ threshold that should result in observation of PBHs. The absence of such an observation rules out this model.}
    \label{fig:pr-constraints}
\end{figure}

The enhanced scalar power spectrum also has key implications for the formation of Primordial Black Holes (PBHs). The parameter space of $\mathcal{P}_{\mathcal{R}}$ inferred from PTA measurements leads to PBH overproduction.
In \Cref{fig:pr-constraints}, we present $\mathcal{P}_{\mathcal{R}}$ derived using the best-fit parameters from our PTA analysis, that results in $\log_{10} \Delta N_{\rm eff} = -3.894$. The spectrum exceeds the PBH production threshold of $\mathcal{P}_{\mathcal{R}} > 10^{-2}$, and leads to PBH overproduction. Hence, it cannot be consistent with current data of PBH abundance.
The compatibility of PBH constraints with $\Delta N_{\rm eff}$ bounds for other SGWB sources, including PGWs \cite{Kumar:2025jfi}, should be examined. We defer this analysis to future work.

Next, in the case of phase transitions, the constraints on the parameters obtained at 68\% CL for the \texttt{Bubble}(\texttt{Sound}) models are $log_{10}\alpha = 0.29^{+0.65}_{-0.29}(0.24^{+0.46}_{-0.56})$, $log_{10} T_* = -0.71^{+0.36}_{-0.47}(-1.71^{+0.38}_{-0.43})$, and $log_{10} H_*R_* = -0.36^{+0.32}_{-0.16}(-0.71^{+0.26}_{-0.43})$. We find $\log_{10} \Delta N_{\rm eff} = -3.28^{+0.28}_{-0.32}$  and $\log_{10} \Delta N_{\rm eff} = -3.53 \pm 0.38$ for the models \texttt{Sound} and \texttt{Bubble} respectively. Since the spectra in both cases follow a broken power law, integrating and extrapolating to 1 $\mu$Hz does not result in any significant change in the value of $\Delta N_{\text{eff}}$, and it remains undetectable, even with SO.

The metastable cosmic strings provide 68\% constraints on the model parameters, ${\rm log_{10}}(G\mu) = -5.44^{+0.76}_{-0.39}$ and $\sqrt{\kappa} = 7.887^{+0.039}_{-0.11}$, whereas a constraint on the derived quantity $\Delta N_{eff}$ integrated until 60 nHz is $\log_{10} \Delta N_{\rm eff} = -1.78^{+0.098}_{-0.210}$.  The extrapolated integral for $f_{max} = \rm 1 \, \mu Hz$ constrains $\Delta N_{eff}$ to be $\log_{10} \Delta N_{eff} = -0.53_{-0.33}^{+0.93}$. Similar to the \texttt{Standard} case this will give us a $>6\sigma$ detection by SO.

\begin{table}[H]
    \centering
    \caption{The inferred 68\% constraints on the derived parameter $\rm log_{10} \Delta N_{eff}$ from NANOGrav 15-year data. For the extrapolated case up to 1 $\mu$Hz, we also show the detectability significance for the present, \texttt{Planck} (TT+TE+EE+lowE+lensing) limit, and for the projected SO limit. The \texttt{Bubble} and \texttt{Sound} cases remain undetectable.}
    \label{tab:neff}
    
    \vspace{0.5em}
    \vspace{1em}
    
    \renewcommand{\arraystretch}{1.3}
    \begin{tabular}{|l|c||c|cc|c|}
        \hline
        \textbf{Model} & \multicolumn{5}{c|}{\textbf{$\log_{10} \Delta N_{\text{eff}}$}} \\
        \cline{2-6}
        & $f_{\text{max}} = 6 \times 10^{-8}$ Hz & $f_{\text{max}} = 1 \times 10^{-6}$ Hz & $\# \sigma_{\textbf{present}}$ & $\# \sigma_{\textbf{SO}}$ & $f_{3 \sigma_{\textbf{SO}}}$ (Hz) \\
        \hline\hline
        
        \texttt{Standard}            & $-2.72^{+0.27}_{-0.32}$ & $-0.54^{+0.66}_{-0.76}$ & 1.52   & 6.41   & $6.65 \times 10^{-7}$ \\
        \texttt{Reheating}           & $-3.1 \pm 1.5$          & $-0.80^{+1.3}_{-1.5}$   & 0.83   & 3.52   & $9.40 \times 10^{-7}$ \\
        \texttt{Running}             & $-2.00 \pm 0.31$        & $-0.13^{+0.86}_{-0.97}$ & 3.90   & 16.4   & $3.15 \times 10^{-7}$ \\
        \texttt{Sound}               & $-3.28^{+0.28}_{-0.32}$ & $-3.26^{+0.28}_{-0.33}$ & --     & --     & -- \\
        \texttt{Bubble}              & $-3.53 \pm 0.38$        & $-3.35^{+0.41}_{-0.49}$ & --     & --     & -- \\
        \texttt{SIGW}                & $-3.894^{+0.031}_{-0.035}$ & $2.336^{+0.231}_{-0.153}$ & $>1000$ & $>1000$ & $2.60 \times 10^{-7}$ \\
        \texttt{Cosmic String}       & $-1.78^{+0.098}_{-0.210}$ & $-0.53^{+0.93}_{-0.33}$ & 1.55   & 6.54   & $3.50 \times 10^{-7}$ \\
        \hline
    \end{tabular} \label{tab: significance}
\end{table}

In \cref{tab: significance}, we summarize the contribution to $\log_{10} \Delta N_{\rm eff}$ integrated up to 60 nHz, corresponding to the current NANOGrav bound, and up to 1 $\mu$Hz, representing a future projected sensitivity, for all the models studied in this work. The third column presents the statistical significance with which these models can be ruled out when their $\log_{10} \Delta N_{\rm eff}$ contributions are compared against current (Planck) and the upcoming SO bounds. In the last column, we report the maximum frequency that a PTA must probe in order to unambiguously rule out or detect a model at greater than 3$\sigma$ confidence. We find that if extrapolated to $\mu$Hz frequencies, the $\log_{10} \Delta N_{\rm eff}$ contributions from the \texttt{Running} and \texttt{SIGW} scenarios are already in strong tension with current bounds, exceeding the 3$\sigma$ level. This makes it unlikely that the NANOGrav signal originates from either of these models.  By contrast, ruling out the remaining scenarios will require PTA experiments to achieve sensitivity at higher frequencies (last column of \cref{tab: significance}). 

\section{Conclusion}

In this study, we have shown that the combination of PTA observations and CMB measurements of extra radiation offers a decisive test of Early-Universe gravitational-wave explanations of the PTA signals. Using Bayesian inference on a variety of well-motivated Early-Universe models—inflation, SIGW, metastable cosmic strings, and first-order phase transitions. 
In the absence of further complication of the models, we find that those models capable of generating a $\Delta N_{\rm eff}$ several orders of magnitude greater than current CMB constraints should be ruled out as explanations for the PTA signal.

Our results are consistent with the previous analysis done in \cite{Wright:2024awr}, which studied phase transitions and the SIGW model; however, unlike our work, that analysis considered a Dirac-delta feature in the primordial scalar spectrum. Moreover, the authors of \cite{Wright:2024awr} generate quasi-random samples uniformly within prior boxes and integrate the spectrum for each sample without evaluating the observational likelihood, so their histograms reflect prior-predictive distributions rather than true posteriors. In contrast, we perform a full Bayesian inference, yielding posterior samples that are weighted by the likelihood and therefore encode both the observational constraints and parameter correlations; $\Delta N_{\rm eff}$ is then evaluated as a derived parameter from these samples, ensuring statistically rigorous credible intervals.

This paper builds directly upon the foundational framework established in the earlier work \cite{Ben-Dayan:2019gll} by moving from general constraints on the scalar-induced tensor power spectrum to targeted, model-specific tests of the Early Universe using combined PTA and CMB observations. Crucially, the new analysis transitions from setting bounds on scalar features to using those bounds as discriminators among competing models. With the exception of the phase transition explanation, all models imply that the PTA signal should be accompanied by a clear measured excess of $\Delta N_{eff}$, or be decisively ruled out. 

In particular, SIGW models and inflationary models with strong running both yield $\Delta N_{\rm eff}$ contributions that are well beyond both Planck limits and the projected sensitivities of near-future CMB surveys such as the Simons Observatory and CMB-S4. These models will be excluded at high confidence (up to $\mathcal{O}(10^3)\sigma$) well before PTA experiments reach $\mu$Hz frequencies. In general PTA measurements up to $\mathcal{O}(few)\times 10^{-7} Hz$ is sufficient for expecting a clear $\Delta N_{\rm eff}$ detection.

Our results highlight the power of joint PTA–CMB analyses to constrain Early-Universe origins of nanohertz-frequency gravitational-wave backgrounds. Any primordial interpretation of the NANOGrav signal would need to be stringently tested for its contribution to $\Delta N_{\rm eff}$, offering an independent and complementary cosmological consistency check. Upcoming improvements in PTA sensitivities, coupled with the precision measurements from CMB experiments, have the clear potential to discriminate between astrophysical and cosmological origins for the observed stochastic background.

\section*{Acknowledgements}
We acknowledge the Ariel HPC Center at Ariel University for providing computing resources that have contributed to the research results reported within this paper. IBD and AV are supported in part by the ``Program of Support of High Energy Physics" Grant by Israeli Council for Higher Education.
\bibliography{ref}

\begin{thebibliography}{100}

\bibitem{NANOGrav:2023gor}
Gabriella Agazie et~al.
\newblock {The NANOGrav 15 yr Data Set: Evidence for a Gravitational-wave
  Background}.
\newblock {\em Astrophys. J. Lett.}, 951(1):L8, 2023.

\bibitem{NANOGrav:2023hde}
Gabriella Agazie et~al.
\newblock {The NANOGrav 15 yr Data Set: Observations and Timing of 68
  Millisecond Pulsars}.
\newblock {\em Astrophys. J. Lett.}, 951(1):L9, 2023.

\bibitem{agazie2023nanograv}
Gabriella Agazie, Akash Anumarlapudi, Anne~M Archibald, Paul~T Baker, Bence
  B{\'e}csy, Laura Blecha, Alexander Bonilla, Adam Brazier, Paul~R Brook, Sarah
  Burke-Spolaor, et~al.
\newblock The nanograv 15-year data set: Constraints on supermassive black hole
  binaries from the gravitational wave background.
\newblock {\em arXiv preprint arXiv:2306.16220}, 2023.

\bibitem{NANOGrav:2023hvm}
Adeela Afzal et~al.
\newblock {The NANOGrav 15 yr Data Set: Search for Signals from New Physics}.
\newblock {\em Astrophys. J. Lett.}, 951(1):L11, 2023.

\bibitem{NANOGrav:2023ctt}
Gabriella Agazie et~al.
\newblock {The NANOGrav 15 yr Data Set: Detector Characterization and Noise
  Budget}.
\newblock {\em Astrophys. J. Lett.}, 951(1):L10, 2023.

\bibitem{johnson2023nanograv}
Aaron~D Johnson, Patrick~M Meyers, Paul~T Baker, Neil~J Cornish, Jeffrey~S
  Hazboun, Tyson~B Littenberg, Joseph~D Romano, Stephen~R Taylor, Michele
  Vallisneri, Sarah~J Vigeland, et~al.
\newblock The nanograv 15-year gravitational-wave background analysis pipeline.
\newblock {\em arXiv preprint arXiv:2306.16223}, 2023.

\bibitem{NANOGrav:2023pdq}
Gabriella Agazie et~al.
\newblock {The NANOGrav 15 yr Data Set: Bayesian Limits on Gravitational Waves
  from Individual Supermassive Black Hole Binaries}.
\newblock {\em Astrophys. J. Lett.}, 951(2):L50, 2023.

\bibitem{Chen:2021rqp}
S.~Chen et~al.
\newblock {Common-red-signal analysis with 24-yr high-precision timing of the
  European Pulsar Timing Array: inferences in the stochastic gravitational-wave
  background search}.
\newblock {\em Mon. Not. Roy. Astron. Soc.}, 508(4):4970--4993, 2021.

\bibitem{Goncharov:2021oub}
Boris Goncharov et~al.
\newblock {On the Evidence for a Common-spectrum Process in the Search for the
  Nanohertz Gravitational-wave Background with the Parkes Pulsar Timing Array}.
\newblock {\em Astrophys. J. Lett.}, 917(2):L19, 2021.

\bibitem{Antoniadis:2022pcn}
J.~Antoniadis et~al.
\newblock {The International Pulsar Timing Array second data release: Search
  for an isotropic gravitational wave background}.
\newblock {\em Mon. Not. Roy. Astron. Soc.}, 510(4):4873--4887, 2022.

\bibitem{NANOGrav:2023hfp}
Gabriella Agazie et~al.
\newblock {The NANOGrav 15-year Data Set: Constraints on Supermassive Black
  Hole Binaries from the Gravitational Wave Background}.
\newblock {\em .}, 6 2023.

\bibitem{Ellis:2023oxs}
John Ellis, Malcolm Fairbairn, Gabriele Franciolini, Gert H\"utsi, Antonio
  Iovino, Marek Lewicki, Martti Raidal, Juan Urrutia, Ville Vaskonen, and Hardi
  Veerm\"ae.
\newblock {What is the source of the PTA GW signal?}
\newblock {\em Phys. Rev. D}, 109(2):023522, 2024.

\bibitem{Ben-Dayan:2023lwd}
Ido Ben-Dayan, Utkarsh Kumar, Udaykrishna Thattarampilly, and Amresh Verma.
\newblock {Probing the early Universe cosmology with NANOGrav: Possibilities
  and limitations}.
\newblock {\em Phys. Rev. D}, 108(10):103507, 2023.

\bibitem{Vagnozzi:2023lwo}
Sunny Vagnozzi.
\newblock {Inflationary interpretation of the stochastic gravitational wave
  background signal detected by pulsar timing array experiments}.
\newblock {\em .}, 6 2023.

\bibitem{Benetti:2021uea}
Micol Benetti, Leila~Lobato Graef, and Sunny Vagnozzi.
\newblock {Primordial gravitational waves from NANOGrav: A broken power-law
  approach}.
\newblock {\em Phys. Rev. D}, 105(4):043520, 2022.

\bibitem{Vagnozzi:2022qmc}
Sunny Vagnozzi and Abraham Loeb.
\newblock {The Challenge of Ruling Out Inflation via the Primordial Graviton
  Background}.
\newblock {\em Astrophys. J. Lett.}, 939(2):L22, 2022.

\bibitem{Das:2023nmm}
Barnali Das, Nur Jaman, and M.~Sami.
\newblock {Gravitational Waves Background (NANOGrav) from Quintessential
  Inflation}.
\newblock {\em .}, 7 2023.

\bibitem{Datta:2023vbs}
Satyabrata Datta.
\newblock {Inflationary gravitational waves, pulsar timing data and
  low-scale-leptogenesis}.
\newblock {\em .}, 7 2023.

\bibitem{Vagnozzi:2020gtf}
Sunny Vagnozzi.
\newblock {Implications of the NANOGrav results for inflation}.
\newblock {\em Mon. Not. Roy. Astron. Soc.}, 502(1):L11--L15, 2021.

\bibitem{Maiti:2024nhv}
Subhasis Maiti, Debaprasad Maity, and L.~Sriramkumar.
\newblock {Constraining inflationary magnetogenesis and reheating via GWs in
  light of PTA data}.
\newblock 1 2024.

\bibitem{Li:2023qua}
Jun-Peng Li, Sai Wang, Zhi-Chao Zhao, and Kazunori Kohri.
\newblock {Primordial non-Gaussianity f $_{NL}$ and anisotropies in
  scalar-induced gravitational waves}.
\newblock {\em JCAP}, 10:056, 2023.

\bibitem{Domenech:2024wao}
Guillem Dom\`enech and Jan Tr\"ankle.
\newblock {From formation to evaporation: Induced gravitational wave probes of
  the primordial black hole reheating scenario}.
\newblock {\em Phys. Rev. D}, 111(6):063528, 2025.

\bibitem{Kohri:2024qpd}
Kazunori Kohri, Takahiro Terada, and Tsutomu~T. Yanagida.
\newblock {Induced gravitational waves probing primordial black hole dark
  matter with the memory burden effect}.
\newblock {\em Phys. Rev. D}, 111(6):063543, 2025.

\bibitem{Chen:2025fcd}
Fei-Yu Chen, Jing-Zhi Zhou, Di~Wu, Zhi-Chao Li, and Peng-Yu Wu.
\newblock {Tensor induced gravitational waves}.
\newblock 7 2025.

\bibitem{Wu:2025gwt}
Di~Wu, Zhi-Chao Li, Peng-Yu Wu, Fei-Yu Chen, and Jing-Zhi Zhou.
\newblock {Probing small-scale primordial power spectra with tensor-scalar
  induced gravitational waves}.
\newblock 7 2025.

\bibitem{Wu:2024qdb}
Di~Wu, Jing-Zhi Zhou, Yu-Ting Kuang, Zhi-Chao Li, Zhe Chang, and Qing-Guo
  Huang.
\newblock {Can tensor-scalar induced GWs dominate PTA observations?}
\newblock {\em JCAP}, 03:045, 2025.

\bibitem{Zhou:2024yke}
Jing-Zhi Zhou, Yu-Ting Kuang, Zhe Chang, and H.~L{\"u}.
\newblock {Constraints on Primordial Black Holes from N$_{eff}$: Scalar-induced
  Gravitational Waves as an Extra Radiation Component}.
\newblock {\em Astrophys. J.}, 979(2):178, 2025.

\bibitem{Zhou:2025djn}
Jing-Zhi Zhou, Zhi-Chao Li, and Di~Wu.
\newblock {Cosmological constraints on small-scale primordial non-Gaussianity}.
\newblock 5 2025.

\bibitem{Chen:2023bms}
Zu-Cheng Chen, Shou-Long Li, Puxun Wu, and Hongwei Yu.
\newblock {NANOGrav hints for first-order confinement-deconfinement phase
  transition in different QCD-matter scenarios}.
\newblock {\em Phys. Rev. D}, 109(4):043022, 2024.

\bibitem{Gouttenoire:2023bqy}
Yann Gouttenoire.
\newblock {First-order Phase Transition interpretation of PTA signal produces
  solar-mass Black Holes}.
\newblock {\em .}, 7 2023.

\bibitem{NANOGrav:2021flc}
Zaven Arzoumanian et~al.
\newblock {Searching for Gravitational Waves from Cosmological Phase
  Transitions with the NANOGrav 12.5-Year Dataset}.
\newblock {\em Phys. Rev. Lett.}, 127(25):251302, 2021.

\bibitem{Bringmann:2023opz}
Torsten Bringmann, Paul~Frederik Depta, Thomas Konstandin, Kai Schmidt-Hoberg,
  and Carlo Tasillo.
\newblock {Does NANOGrav observe a dark sector phase transition?}
\newblock {\em JCAP}, 11:053, 2023.

\bibitem{Bagherian:2025puf}
Hengameh Bagherian, Majid Ekhterachian, and Stefan Stelzl.
\newblock {The Bearable Inhomogeneity of the Baryon Asymmetry}.
\newblock 5 2025.

\bibitem{Guo:2023hyp}
Shu-Yuan Guo, Maxim Khlopov, Xuewen Liu, Lei Wu, Yongcheng Wu, and Bin Zhu.
\newblock {Footprints of axion-like particle in pulsar timing array data and
  James Webb Space Telescope observations}.
\newblock {\em Sci. China Phys. Mech. Astron.}, 67(11):111011, 2024.

\bibitem{Datta:2024bqp}
Satyabrata Datta and Rome Samanta.
\newblock {Cosmic superstrings, metastable strings and ultralight primordial
  black holes: from NANOGrav to LIGO and beyond}.
\newblock {\em JHEP}, 02:095, 2025.

\bibitem{Blasi:2020mfx}
Simone Blasi, Vedran Brdar, and Kai Schmitz.
\newblock {Has NANOGrav found first evidence for cosmic strings?}
\newblock {\em Phys. Rev. Lett.}, 126(4):041305, 2021.

\bibitem{Kume:2024adn}
Jun'ya Kume and Mark Hindmarsh.
\newblock {Revised bounds on local cosmic strings from NANOGrav observations}.
\newblock {\em JCAP}, 12:001, 2024.

\bibitem{Hindmarsh:2022awe}
Mark Hindmarsh and Jun'ya Kume.
\newblock {Multi-messenger constraints on Abelian-Higgs cosmic string
  networks}.
\newblock {\em JCAP}, 04:045, 2023.

\bibitem{Brandenberger:2011et}
Robert~H. Brandenberger.
\newblock {String Gas Cosmology: Progress and Problems}.
\newblock {\em Class. Quant. Grav.}, 28:204005, 2011.

\bibitem{Ben-Dayan:2016iks}
Ido Ben-Dayan.
\newblock {Gravitational Waves in Bouncing Cosmologies from Gauge Field
  Production}.
\newblock {\em JCAP}, 09:017, 2016.

\bibitem{Ben-Dayan:2018ksd}
Ido Ben-Dayan and Judy Kupferman.
\newblock {Sourced scalar fluctuations in bouncing cosmology}.
\newblock {\em JCAP}, 07:050, 2019.
\newblock [Erratum: JCAP 12, E01 (2020)].

\bibitem{Brandenberger:2020tcr}
Robert Brandenberger and Ziwei Wang.
\newblock {Nonsingular Ekpyrotic Cosmology with a Nearly Scale-Invariant
  Spectrum of Cosmological Perturbations and Gravitational Waves}.
\newblock {\em Phys. Rev. D}, 101(6):063522, 2020.

\bibitem{Ben-Dayan:2024aec}
Ido Ben-Dayan, Gianluca Calcagni, Maurizio Gasperini, Anupam Mazumdar, Eliseo
  Pavone, Udaykrishna Thattarampilly, and Amresh Verma.
\newblock {Gravitational-wave background in bouncing models from
  semi-classical, quantum and string gravity}.
\newblock {\em JCAP}, 09:058, 2024.

\bibitem{Ben-Dayan:2019gll}
Ido Ben-Dayan, Brian Keating, David Leon, and Ira Wolfson.
\newblock {Constraints on scalar and tensor spectra from $N_{eff}$}.
\newblock {\em JCAP}, 06:007, 2019.

\bibitem{Jiang:2023gfe}
Jun-Qian Jiang, Yong Cai, Gen Ye, and Yun-Song Piao.
\newblock {Broken blue-tilted inflationary gravitational waves: a joint
  analysis of NANOGrav 15-year and BICEP/Keck 2018 data}.
\newblock {\em JCAP}, 05:004, 2024.

\bibitem{Hou_2013}
Zhen Hou, Ryan Keisler, Lloyd Knox, Marius Millea, and Christian Reichardt.
\newblock How massless neutrinos affect the cosmic microwave background damping
  tail.
\newblock {\em Physical Review D}, 87(8), April 2013.

\bibitem{Abazajian:2012ys}
K.~N. Abazajian et~al.
\newblock {Light Sterile Neutrinos: A White Paper}.
\newblock 4 2012.

\bibitem{SO:2024ntl}
Nicholas Galitzki et~al.
\newblock {The Simons Observatory: Design, Integration, and Testing of the
  Small Aperture Telescopes}.
\newblock {\em Astrophys. J. Suppl.}, 274(2):33, 2024.

\bibitem{CMB-S4:2022pft}
Patricio~A. Gallardo et~al.
\newblock {Optical design concept of the CMB-S4 large-aperture telescopes and
  cameras}.
\newblock {\em Proc. SPIE Int. Soc. Opt. Eng.}, 12190:189, 2022.

\bibitem{LiteBIRD:2022cnt}
E.~Allys et~al.
\newblock {Probing Cosmic Inflation with the LiteBIRD Cosmic Microwave
  Background Polarization Survey}.
\newblock {\em PTEP}, 2023(4):042F01, 2023.

\bibitem{Planck:2018vyg}
N.~Aghanim et~al.
\newblock {Planck 2018 results. VI. Cosmological parameters}.
\newblock {\em Astron. Astrophys.}, 641:A6, 2020.
\newblock [Erratum: Astron.Astrophys. 652, C4 (2021)].

\bibitem{Nakayama:2018ptw}
Kazunori Nakayama and Yong Tang.
\newblock {Stochastic Gravitational Waves from Particle Origin}.
\newblock {\em Phys. Lett. B}, 788:341--346, 2019.

\bibitem{NANOGrav:2023tcn}
Gabriella Agazie et~al.
\newblock {The NANOGrav 15-year Data Set: Search for Anisotropy in the
  Gravitational-Wave Background}.
\newblock {\em .}, 6 2023.

\bibitem{Brazier:2019mmu}
A.~Brazier et~al.
\newblock {The NANOGrav Program for Gravitational Waves and Fundamental
  Physics}.
\newblock 8 2019.

\bibitem{Rosado:2015epa}
Pablo~A. Rosado, Alberto Sesana, and Jonathan Gair.
\newblock {Expected properties of the first gravitational wave signal detected
  with pulsar timing arrays}.
\newblock {\em Mon. Not. Roy. Astron. Soc.}, 451(3):2417--2433, 2015.

\bibitem{Moore:2014eua}
Christopher~J. Moore, Stephen~R. Taylor, and Jonathan~R. Gair.
\newblock {Estimating the sensitivity of pulsar timing arrays}.
\newblock {\em Class. Quant. Grav.}, 32(5):055004, 2015.

\bibitem{Bailes:2021tot}
M.~Bailes et~al.
\newblock {Gravitational-wave physics and astronomy in the 2020s and 2030s}.
\newblock {\em Nature Rev. Phys.}, 3(5):344--366, 2021.

\bibitem{Perera:2018pts}
B.~B.~P. Perera et~al.
\newblock {Improving timing sensitivity in the microhertz frequency regime:
  limits from PSR J1713$+$0747 on gravitational waves produced by super-massive
  black-hole binaries}.
\newblock {\em Mon. Not. Roy. Astron. Soc.}, 478(1):218--227, 2018.

\bibitem{Boyle:2007zx}
Latham~A. Boyle and Alessandra Buonanno.
\newblock {Relating gravitational wave constraints from primordial
  nucleosynthesis, pulsar timing, laser interferometers, and the CMB:
  Implications for the early Universe}.
\newblock {\em Phys. Rev. D}, 78:043531, 2008.

\bibitem{Ben-Dayan:2009fyj}
Ido Ben-Dayan and Ram Brustein.
\newblock {Cosmic Microwave Background Observables of Small Field Models of
  Inflation}.
\newblock {\em JCAP}, 09:007, 2010.

\bibitem{Bringmann:2011ut}
Torsten Bringmann, Pat Scott, and Yashar Akrami.
\newblock {Improved constraints on the primordial power spectrum at small
  scales from ultracompact minihalos}.
\newblock {\em Phys. Rev. D}, 85:125027, 2012.

\bibitem{Chluba:2012we}
Jens Chluba, Adrienne~L. Erickcek, and Ido Ben-Dayan.
\newblock {Probing the inflaton: Small-scale power spectrum constraints from
  measurements of the CMB energy spectrum}.
\newblock {\em Astrophys. J.}, 758:76, 2012.

\bibitem{Ben-Dayan:2013eza}
Ido Ben-Dayan and Tigran Kalaydzhyan.
\newblock {Constraining the primordial power spectrum from SNIa lensing
  dispersion}.
\newblock {\em Phys. Rev. D}, 90(8):083509, 2014.

\bibitem{Ben-Dayan:2014iya}
Ido Ben-Dayan.
\newblock {Lensing dispersion of SNIa and small scales of the primordial power
  spectrum}.
\newblock In {\em {49th Rencontres de Moriond on Cosmology}}, pages 95--100,
  2014.

\bibitem{Ben-Dayan:2015zha}
Ido Ben-Dayan and Ryuichi Takahashi.
\newblock {Constraints on small-scale cosmological fluctuations from SNe
  lensing dispersion}.
\newblock {\em Mon. Not. Roy. Astron. Soc.}, 455(1):552--562, 2016.

\bibitem{Ben-Dayan:2013fva}
Ido Ben-Dayan, Shenglin Jing, Alexander Westphal, and Clemens Wieck.
\newblock {Accidental inflation from K{\"a}hler uplifting}.
\newblock {\em JCAP}, 03:054, 2014.

\bibitem{Artymowski:2020pci}
Micha\l{} Artymowski, Ido Ben-Dayan, and Udaykrishna Thattarampilly.
\newblock {Sourced fluctuations in generic slow contraction}.
\newblock {\em JCAP}, 06:010, 2021.

\bibitem{Ben-Dayan:2023rlj}
Ido Ben-Dayan and Udaykrishna Thattarampilly.
\newblock {Requiem to {\textquotedblleft}proof of inflation{\textquotedblright}
  or sourced fluctuations in a non-singular bounce}.
\newblock {\em JCAP}, 06:004, 2024.

\bibitem{Mollerach:2003nq}
Silvia Mollerach, Diego Harari, and Sabino Matarrese.
\newblock {CMB polarization from secondary vector and tensor modes}.
\newblock {\em Phys. Rev. D}, 69:063002, 2004.

\bibitem{Baumann:2007zm}
Daniel Baumann, Paul~J. Steinhardt, Keitaro Takahashi, and Kiyotomo Ichiki.
\newblock {Gravitational Wave Spectrum Induced by Primordial Scalar
  Perturbations}.
\newblock {\em Phys. Rev. D}, 76:084019, 2007.

\bibitem{Ananda:2006af}
Kishore~N. Ananda, Chris Clarkson, and David Wands.
\newblock {The Cosmological gravitational wave background from primordial
  density perturbations}.
\newblock {\em Phys. Rev. D}, 75:123518, 2007.

\bibitem{Kumar:2024bvp}
Utkarsh Kumar, Udaykrishna Thattarampilly, and Pankaj Chaturvedi.
\newblock {Probe of spatial geometry from scalar induced gravitational waves}.
\newblock 10 2024.

\bibitem{Domenech:2021ztg}
Guillem Dom\`enech.
\newblock {Scalar Induced Gravitational Waves Review}.
\newblock {\em Universe}, 7(11):398, 2021.

\bibitem{Pi:2020otn}
Shi Pi and Misao Sasaki.
\newblock {Gravitational Waves Induced by Scalar Perturbations with a Lognormal
  Peak}.
\newblock {\em JCAP}, 09:037, 2020.

\bibitem{Espinosa:2018eve}
Jos\'e~Ram\'on Espinosa, Davide Racco, and Antonio Riotto.
\newblock {A Cosmological Signature of the SM Higgs Instability: Gravitational
  Waves}.
\newblock {\em JCAP}, 09:012, 2018.

\bibitem{Kohri:2018awv}
Kazunori Kohri and Takahiro Terada.
\newblock {Semianalytic calculation of gravitational wave spectrum nonlinearly
  induced from primordial curvature perturbations}.
\newblock {\em Phys. Rev. D}, 97(12):123532, 2018.

\bibitem{Barger:2007im}
Vernon Barger, Paul Langacker, Mathew McCaskey, Michael~J. Ramsey-Musolf, and
  Gabe Shaughnessy.
\newblock {LHC Phenomenology of an Extended Standard Model with a Real Scalar
  Singlet}.
\newblock {\em Phys. Rev. D}, 77:035005, 2008.

\bibitem{Damgaard:2015con}
P.~H. Damgaard, A.~Haarr, D.~O'Connell, and A.~Tranberg.
\newblock {Effective Field Theory and Electroweak Baryogenesis in the
  Singlet-Extended Standard Model}.
\newblock {\em JHEP}, 02:107, 2016.

\bibitem{Cline:1996mga}
James~M. Cline and Pierre-Anthony Lemieux.
\newblock {Electroweak phase transition in two Higgs doublet models}.
\newblock {\em Phys. Rev. D}, 55:3873--3881, 1997.

\bibitem{FileviezPerez:2008bj}
Pavel Fileviez~Perez, Hiren~H. Patel, Michael.~J. Ramsey-Musolf, and Kai Wang.
\newblock {Triplet Scalars and Dark Matter at the LHC}.
\newblock {\em Phys. Rev. D}, 79:055024, 2009.

\bibitem{Dorsch:2013wja}
G.~C. Dorsch, S.~J. Huber, and J.~M. No.
\newblock {A strong electroweak phase transition in the 2HDM after LHC8}.
\newblock {\em JHEP}, 10:029, 2013.

\bibitem{Lewicki:2020azd}
Marek Lewicki and Ville Vaskonen.
\newblock {Gravitational waves from colliding vacuum bubbles in gauge
  theories}.
\newblock {\em Eur. Phys. J. C}, 81(5):437, 2021.
\newblock [Erratum: Eur.Phys.J.C 81, 1077 (2021)].

\bibitem{Cutting:2018tjt}
Daniel Cutting, Mark Hindmarsh, and David~J. Weir.
\newblock {Gravitational waves from vacuum first-order phase transitions: from
  the envelope to the lattice}.
\newblock {\em Phys. Rev. D}, 97(12):123513, 2018.

\bibitem{Ellis:2018mja}
John Ellis, Marek Lewicki, and Jos{\'e}~Miguel No.
\newblock {On the Maximal Strength of a First-Order Electroweak Phase
  Transition and its Gravitational Wave Signal}.
\newblock {\em JCAP}, 04:003, 2019.

\bibitem{Ellis:2020awk}
John Ellis, Marek Lewicki, and Jos{\'e}~Miguel No.
\newblock {Gravitational waves from first-order cosmological phase transitions:
  lifetime of the sound wave source}.
\newblock {\em JCAP}, 07:050, 2020.

\bibitem{Huang:2017laj}
Fa~Peng Huang and Xinmin Zhang.
\newblock {Probing the gauge symmetry breaking of the early universe in 3-3-1
  models and beyond by gravitational waves}.
\newblock {\em Phys. Lett. B}, 788:288--294, 2019.

\bibitem{Hindmarsh:2017gnf}
Mark Hindmarsh, Stephan~J. Huber, Kari Rummukainen, and David~J. Weir.
\newblock {Shape of the acoustic gravitational wave power spectrum from a first
  order phase transition}.
\newblock {\em Phys. Rev. D}, 96(10):103520, 2017.
\newblock [Erratum: Phys.Rev.D 101, 089902 (2020)].

\bibitem{Espinosa:2010hh}
Jose~R. Espinosa, Thomas Konstandin, Jose~M. No, and Geraldine Servant.
\newblock {Energy Budget of Cosmological First-order Phase Transitions}.
\newblock {\em JCAP}, 06:028, 2010.

\bibitem{Guo:2020grp}
Huai-Ke Guo, Kuver Sinha, Daniel Vagie, and Graham White.
\newblock {Phase Transitions in an Expanding Universe: Stochastic Gravitational
  Waves in Standard and Non-Standard Histories}.
\newblock {\em JCAP}, 01:001, 2021.

\bibitem{Weir:2017wfa}
David~J. Weir.
\newblock {Gravitational waves from a first order electroweak phase transition:
  a brief review}.
\newblock {\em Phil. Trans. Roy. Soc. Lond. A}, 376(2114):20170126, 2018.
\newblock [Erratum: Phil.Trans.Roy.Soc.Lond.A 381, 20230212 (2023)].

\bibitem{Jinno:2016vai}
Ryusuke Jinno and Masahiro Takimoto.
\newblock {Gravitational waves from bubble collisions: An analytic derivation}.
\newblock {\em Phys. Rev. D}, 95(2):024009, 2017.

\bibitem{Hindmarsh:2020hop}
Mark~B. Hindmarsh, Marvin L{\"u}ben, Johannes Lumma, and Martin Pauly.
\newblock {Phase transitions in the early universe}.
\newblock {\em SciPost Phys. Lect. Notes}, 24:1, 2021.

\bibitem{Cheng:2024axj}
Hanyu Cheng and Luca Visinelli.
\newblock {Future targets for light gauge bosons from cosmic strings}.
\newblock {\em Phys. Dark Univ.}, 46:101667, 2024.

\bibitem{Vilenkin:2000jqa}
A.~Vilenkin and E.~P.~S. Shellard.
\newblock {\em {Cosmic Strings and Other Topological Defects}}.
\newblock Cambridge University Press, 7 2000.

\bibitem{Dvali:1994ms}
G.~R. Dvali, Q.~Shafi, and Robert~K. Schaefer.
\newblock {Large scale structure and supersymmetric inflation without fine
  tuning}.
\newblock {\em Phys. Rev. Lett.}, 73:1886--1889, 1994.

\bibitem{Jeannerot:2003qv}
Rachel Jeannerot, Jonathan Rocher, and Mairi Sakellariadou.
\newblock {How generic is cosmic string formation in SUSY GUTs}.
\newblock {\em Phys. Rev. D}, 68:103514, 2003.

\bibitem{Vilenkin:1982hm}
A.~Vilenkin.
\newblock {COSMOLOGICAL EVOLUTION OF MONOPOLES CONNECTED BY STRINGS}.
\newblock {\em Nucl. Phys. B}, 196:240--258, 1982.

\bibitem{Preskill:1992ck}
John Preskill and Alexander Vilenkin.
\newblock {Decay of metastable topological defects}.
\newblock {\em Phys. Rev. D}, 47:2324--2342, 1993.

\bibitem{Leblond:2009fq}
Louis Leblond, Benjamin Shlaer, and Xavier Siemens.
\newblock {Gravitational Waves from Broken Cosmic Strings: The Bursts and the
  Beads}.
\newblock {\em Phys. Rev. D}, 79:123519, 2009.

\bibitem{Monin:2008mp}
A.~Monin and M.~B. Voloshin.
\newblock {The Spontaneous breaking of a metastable string}.
\newblock {\em Phys. Rev. D}, 78:065048, 2008.

\bibitem{Buchmuller:2021mbb}
Wilfried Buchmuller, Valerie Domcke, and Kai Schmitz.
\newblock {Stochastic gravitational-wave background from metastable cosmic
  strings}.
\newblock {\em JCAP}, 12(12):006, 2021.

\bibitem{Maggiore:1999vm}
Michele Maggiore.
\newblock {Gravitational wave experiments and early universe cosmology}.
\newblock {\em Phys. Rept.}, 331:283--367, 2000.

\bibitem{Mitridate:2023oar}
Andrea Mitridate, David Wright, Richard von Eckardstein, Tobias Schr\"oder,
  Jonathan Nay, Ken Olum, Kai Schmitz, and Tanner Trickle.
\newblock {PTArcade}.
\newblock 6 2023.

\bibitem{Lamb:2023jls}
William~G. Lamb, Stephen~R. Taylor, and Rutger van Haasteren.
\newblock {Rapid refitting techniques for Bayesian spectral characterization of
  the gravitational wave background using pulsar timing arrays}.
\newblock {\em Phys. Rev. D}, 108(10):103019, 2023.

\bibitem{Planck:2018jri}
Y.~Akrami et~al.
\newblock {Planck 2018 results. X. Constraints on inflation}.
\newblock {\em Astron. Astrophys.}, 641:A10, 2020.

\bibitem{Kumar:2025jfi}
Utkarsh Kumar.
\newblock {Primordial Gravitational Wave Background as a Probe of the
  Primordial Black Holes}.
\newblock 7 2025.

\bibitem{Wright:2024awr}
David Wright, John~T. Giblin, and Jeffrey Hazboun.
\newblock {CMB and energy conservation limits on nanohertz gravitational
  waves}.
\newblock 9 2024.

\end{thebibliography}
\end{document}